\documentclass[prc,showpacs,groupedaddress,floatfix,twocolumn]{revtex4}
\usepackage{graphicx}
\usepackage{dcolumn}
\usepackage{bm}
\usepackage{amsmath}
\usepackage{amsfonts}
\usepackage{amssymb}
\usepackage{amsmath}
\usepackage{amsfonts}
\usepackage{amssymb}
\usepackage{graphicx}
\usepackage{epstopdf}
\usepackage{color}%
\providecommand{\U}[1]{\protect\rule{.1in}{.1in}}

\newcommand{\rra}{\mathbf{r}_{1}}
\newcommand{\rrb}{\mathbf{r}_{2}}

\begin{document}
\title{Large-scale Continuum random-phase approximation predictions of dipole strength for astrophysical applications}
\author{I. Daoutidis\footnote{Electronic address: idaoutid@ulb.ac.be} }{
}
\author{S. Goriely\footnote{Electronic address: sgoriely@ulb.ac.be}}{
}

\affiliation{Institut d'Astronomie et d'Astrophysique, Universit\'e Libre de Bruxelles, B-1050, Belgium}
\date{\today}

\begin{abstract}
Large-scale calculations of the $E1$ strength are performed within the random phase approximation (RPA) based on the relativistic point-coupling mean
field approach in order to derive the radiative neutron capture cross sections for all nuclei of astrophysical interest. While the coupling to the
single-particle continuum is taken into account in an explicit and self-consistent way, additional corrections like the coupling to complex
configurations and the temperature and deformation effects are included in a phenomenological way to account for a complete description of the nuclear
dynamical problem. It is shown that the resulting $E1$-strength function based on the PCF1 force is in close agreement with photoabsorption data as
well as the available experimental $E1$ strength data at low energies. For neutron-rich nuclei, as well as light neutron-deficient nuclei, a low-lying
so-called pygmy resonance is found systematically in the 5--10~MeV region. The corresponding strength can reach 10\% of the giant dipole strength in
the neutron-rich region and about 5\% in the neutron-deficient region, and is found to be reduced in the vicinity of the shell closures. Finally, the
neutron capture reaction rates of neutron-rich nuclei is found to be about 2--5 times larger than those predicted on the basis of the nonrelativistic
RPA calculation and about a factor 50 larger than obtained with traditional Lorentzian-type approaches.
\end{abstract}

\maketitle

\section{Introduction}

The investigation of the isovector giant dipole resonance (GDR) is one of the fundamental problems in nuclear physics and astrophysics. These
collective excitations can be studied experimentally by photoabsorption processes and provide essential nuclear inputs for the understanding of
reaction mechanisms taking place during the synthesis of chemical elements in stars.

Despite new successes of recent large accelerator facilities, the information we get on the GDR properties is still limited to about hundred nuclei in
the valley of $\beta$ stability. All the remaining few thousands exotic proton- or neutron-rich nuclei can be accessed by nuclear theory only. The aim
is therefore to provide the necessary theoretical tools towards a universal description of the nuclear structure phenomena, up to the limit of the
nuclear stability.

Over the years, a lot of effort has been devoted to improve the theoretical description of the main collective properties, such as the $\gamma$-ray
strength function. In particular, the random-phase approximation (RPA), based on the relativistic mean field (RMF) theory, has proven to be a robust
theory \cite{HG.89,DF.90,MTG.97}. Pairing correlations for the study of open-shell nuclei as well as the self-consistent description of the residual
interaction \cite{RMG.01} have also been included over the years.

However, a complete and coherent description of the dynamical properties of the nuclei over the entire nuclear chart remains an open problem and
additional corrections beyond the conventional RPA problem are necessary, as discussed the present work.

At first, the coupling to the positive-energy continuum has to be taken into account explicitly. This is important,because the alternative approach,
namely the discretization of the continuum, though simpler, requires a basis truncation, which eventually leads to a mixing between spurious and
physical states of the excitation spectrum. In addition, this discretized RPA includes a large amount of states which increases considerably the
numerical effort.

In fact, the exact treatment of the continuum using the Green's function representation not only can solve these two problems simultaneously, but
also allows for a direct calculation of the escape width of the giant resonance. Although this so-called continuum RPA (CRPA) is known to be a
powerful approach, it has been rarely applied in the past \cite{LG.76b,SB.75,KLL.98,HS.01,Mat.01}. Only recently has a continuum QRPA approach been
successfully developed within a relativistic framework to study giant multipole resonances for spherical nuclei \cite{DR.09,DR.11}.

Furthermore, many applications, such as nuclear astrophysics, require the estimate of the $\gamma$-ray strength function for deformed nuclei.
Recently, microscopic calculations with axially deformed RPA approaches \cite{PKR.09,PR.08,MPD.11} have been able to study medium and heavy deformed
nuclei showing a considerable success in the prediction of the GDR properties. In particular they confirm an important effect, namely the splitting of
the resonance peak into two components. 

Unfortunately, the axially deformed RPA faces several difficulties, mostly related to the numerical effort and the isolation of the spurious state.
For this reason the deformation effect is still treated in all large-scale calculations in a phenomenological way. A similar approach is followed in
the present work.

The CRPA approach \cite{DR.11} is developed in such a way that only one-particle one-hole (1p1h) configurations are taken into account, so that the
coupling to more complex configurations, such as 2p2h, is not described in an explicit way. These couplings can be taken into account in the framework
of the relativistic or nonrelativistic time-blocking approximation \cite{LPT.10,LRT.09,AGKK.11}. However, such a method cannot be applied to
large-scale calculations yet, essentially due to the numerical effort it represents, so that alternative phenomenological approximations need to be
considered \cite{DNSW.90,SW.88}. A similar approach is followed in the present work, as described in Sec.~\ref{CORRECT}.

Regarding calculations of astrophysical interest, it is of importance to describe not only the photoabsorption processes, but also the reverse
radiative particle-capture processes. Reaction theory relates the $\gamma$-transmission coefficient for excited states to the ground state
photoabsorption assuming the giant resonance to be built on each excited state. The description of the photo-deexcitation requires however the
introduction of a temperature dependence due to collisions between quasiparticles. As traditionally done, a temperature-dependent GDR width can be
applied and has been shown to improve the predictions \cite{Khan.04,LG.10}.

Finally, as far as pairing correlations are concerned, it has been extensively discussed in the past that the Hartree-Fock-Bogoliubov (HFB)
\cite{DFT.84,DNW.96} or the relativistic Hartree-Bogoliubov (RHB) approximation \cite{Vretenar05} give a reliable description of both the ground state
and the collective excitations of open shell nuclei. A proper treatment of the pairing correlations is even more important in nuclei close to the
drip lines where the pairing gap becomes of the order of the particle emission threshold. However, this approach is not yet available within the
relativistic CRPA \cite{DR.11}, so that open shell nuclei are still treated within the BCS model applied in the mean field as well as the dynamical
system. Until an RHB plus CRPA is developed, the BCS model can be regarded as a useful approximation of the pairing
effect on the $\gamma$-ray strength function.

The purpose of the present paper is to determine systematically the CRPA $E1$ strength function for the entire nuclear chart within the relativistic
mean field approach. For practical applications, all the above mentioned corrections (correct description of the continuum, spreading width,
deformation effects, temperature effects) are included for a proper prediction of the reaction cross section. The paper is organized as follows. In
Sec.~\ref{CRPA} we briefly describe the relativistic quasiparticle RPA, where the continuum is explicitly included. In Sec.~\ref{CORRECT} the
damping of the collective motion, the impact of deformation on the GDR and the temperature dependence are included in the model and the new strength
is compared with experimental data. In Sec.~\ref{HUGE} a large-scale calculation for about 8000 nuclei is shown to give interesting properties of
the nuclear collective motion. The proton pygmy resonance of neutron deficient nuclei is extensively studied in Sec.~\ref{PPDR}. Finally, in
Sec.~\ref{RATES} we use the final strength to study the radiative neutron capture rates for all the 8000 nuclei and report our conclusions in
Sec.~\ref{SUMMARY}.

\section{Relativistic Continuum QRPA}
\label{CRPA}
As in all the relativistic models, the nucleons are described as point like Dirac particles, which move inside a nuclear mean field. This mean field
is phenomenological and can be described by the exchange of effective mesons, as is done in the Walecka model~\cite{Wal.74} or by using the zero range
or point coupling terms. This latter description of the RMF introduced in the early 1990s \cite{MM.89,HMM.94} has since been improved
substantially
and is used in our work to describe the ground state and collective properties. This approach has the main advantage of simplifying considerably the
numerical calculations without loosing in predictive power as compared to the meson exchange mean field.

In this work, we use the point coupling Lagrangian PCF1~\cite{BMM.02} expanded in powers of the nucleon scalar (S), vector (V) and isovector-vector
(TV) fields:

\begin{eqnarray}
\label{Lagrangian}
 \mathcal{L}&=&\bar{\psi}(i\gamma\cdot\partial-m)\psi \nonumber \\
&-&\sum_{c}\frac{1}{2}\alpha_{c}[\rho](\bar{\psi}\hat{\Gamma}_{c}\psi)(\bar{\psi}\hat{\Gamma}_{c}\psi)
-\frac{1}{2}\delta_{c}(\partial_{\nu}\bar{\psi}\hat{\Gamma}_{c}\psi)(\partial^{\nu}\bar{\psi}\hat{\Gamma}_{c}\psi)\nonumber\\
&-& e\bar{\psi}\gamma\cdot\frac{(1-\tau_{3})}{2}\psi,\qquad\qquad\qquad(c=S,V,TV)
\end{eqnarray}
where the Dirac vertices $\hat{\Gamma}_{c}$ have the explicit form
\begin{equation}
 \hat{\Gamma}_{\text{S}} = 1\qquad \hat{\Gamma}_{\text{V}} = \gamma_{\mu} \qquad \hat{\Gamma}_{\text{TV}} = \gamma_{\mu}\vec{\tau}.
\end{equation}
For the PCF1 force, the index $c$ takes just seven values, namely one scalar S, three isoscalars V and three isovector-vectors TV. The gradient terms
$\delta_{c}(\partial_{\nu}\bar{\psi}\hat{\Gamma}_{c}\psi)$ in the Lagrangian $\mathcal{L}$ are used to simulate the finite range of the nuclear
interaction and consequently improve the description of the nuclear surface properties. The energy functional deduced from Eq.~(\ref{Lagrangian}) is
sufficient to describe the nuclear ground state, the single-particle Hamiltonian being given by:
\begin{equation}
 h_{\alpha \beta}=\frac{\delta E[\hat{\rho}]}{\delta\hat{\rho}_{\alpha\beta}}.
\end{equation}

To describe the nuclear response to an external field $F$ induced by a particle or by photoabsorption, the strength function

\begin{equation}
\label{strenghtfunction}
S(E)=-\frac{1}{\pi}\text{Im}\sum_{\alpha\beta\alpha^{\prime}\beta^{\prime}} F_{\alpha\beta}^{\ast} R_{\alpha\beta\alpha^{\prime}\beta^{\prime}}
(E)F_{\alpha^{\prime}\beta^{\prime}}
\end{equation}
needs to be estimated. The Greek indices $\alpha,\beta$ indicate the various degrees of freedom of a nucleon ($\mathbf{r},s,t,d$), where $s$ is the
spin, $t$ the isospin coordinate, and $d=1,2$ labels the large and small components. If we consider the case of a weak external field and thus a small
amplitude variation of the static solution, the response function $R_{\alpha\beta\alpha^{\prime}\beta^{\prime}}(E)$ can be deduced from the
\textit{linearized Bethe-Salpeter} equation
\begin{eqnarray}
\label{Bethe-Salpeter}
R_{\alpha\beta\alpha^{\prime}\beta^{\prime}}(E) &=&
R_{\alpha\beta\alpha^{\prime}\beta^{\prime}}^{\,0}(E) \\ &+& \sum_{\gamma\delta\gamma^{\prime}\delta^{\prime}} \,\,
R^0_{\alpha\beta\gamma^{\prime}\delta{\prime}}(E) V_{\gamma\delta\gamma^{\prime}\delta^{\prime}}^{\text{ph}}
R_{\gamma^{\prime}\delta^{\prime}\alpha^{\prime}\beta^{\prime}}(E).\nonumber
\end{eqnarray}
The residual interaction $V_{\gamma\delta\gamma^{\prime}\delta^{\prime}}^{\text{ph}}$ of Eq.~(\ref{Bethe-Salpeter}) is connected to the static problem
via the second derivative of the energy functional
\begin{equation}
\label{energy_variation}
V_{\gamma\delta\gamma^{\prime}\delta^{\prime}}^{\text{ph}} = \frac{\delta^{2}E[\hat{\rho}]}
{\delta\hat{\rho}_{\gamma\delta}\delta\hat{\rho}_{\gamma^{\prime}\delta^{\prime}}}.
\end{equation}
This expression of $V_{\gamma\delta\gamma^{\prime}\delta^{\prime}}^{\text{ph}}$ includes in addition to the linear time-like fields, the space-like
terms, the Coulomb part and the rearrangement terms, as a result of the second derivation \cite{DR.11}. In this way, a self-consistent solution is
ensured, which is of particular importance for the description of exotic nuclei for which no data exists.

In coordinate representation, the indices $\alpha$,$\beta,\ldots$ in Eq.~(\ref{Bethe-Salpeter}) are abbreviations for the ``coordinates'' $1=({\bm
r}_1,s_1,t_1,d_1)$. The simplicity of the point coupling Lagrangian allows us to express the effective interaction of Eq.~(\ref{energy_variation}) as
a sum of separable terms, i.e
\begin{equation}\label{separable}
V^{\text{ph}}(1,2)=\sum_{c} Q_{c}^{*(1)} (r) \upsilon_{c}(r)Q_{c}^{(2)} (r)
\end{equation} 
with the local channel operators $Q_{c}^{(1)}({r)}$ defined by
\begin{equation}
\label{channel-operator}
\hat{Q}_{c}^{(1)}(r)=(-)^{S_{c}}\frac{\delta(r-r_1)}{rr_1} \hat{\Gamma}^{(1)}_{c} Y_L (\Omega_{1}).
\end{equation}
These single-particle operators are characterized by the channel index defined in the spin-isospin space ($c$) as well as the coordinate space ($r$).
Therefore, assuming a coordinate mesh $r$ of 50 points and having $c=7$, the size of the interaction matrix is typically smaller than $350\times\,350$
which significantly reduces the numerical effort.

Finally, by inserting the effective interaction of Eq.~(\ref{separable}) into the Bethe-Salpeter equation (\ref{Bethe-Salpeter}), we get the
expression
\begin{equation}
\mathcal{R}_{cc^{\prime}}(E)=
\mathcal{R}_{cc^{\prime}}^{\,0}(E)+
\sum_{c^{\prime\prime}}\mathcal{R}_{cc^{\prime\prime}}^{\,0}(E)\upsilon_{c^{\prime\prime}}(r)\mathcal{R}_{c^{\prime\prime}c^{\prime}}(E),
\label{reduced-bethe-salpeter}
\end{equation}
where the reduced response function is given by
\begin{equation}
\mathcal{R}_{cc^{\prime}}(E)=Q_{c}^{*(1)}R(E)Q_{c^{\prime}}^{(2)}.
\label{reduced-response}
\end{equation}
Equation (\ref{reduced-bethe-salpeter}) has the same formal solution as Eq.~(\ref{Bethe-Salpeter}), but is simpler to solve. At the end, one has
to calculate the free response function $\mathcal{R}_{cc^{\prime}}^{\,0}(E)$ which corresponds to the basic quantity of the CRPA approach.

According to \cite{DR.09,DR.11}, in order to treat the coupling to the continuum exactly, the nonspectral representation $\mathcal{R}_{cont}^{0}$
needs to be estimated from 
\begin{eqnarray}
\label{cont_term}
\mathcal{R}_{cont}^{0} &=& \sum_{k}v_{k}^{2}\langle k(r)|Q_{c}g(r,r^{\prime};E-E_{k}+\lambda) \\ 
&&+g(r^{\prime},r;-E-E_{k}+\lambda)Q_{c^{\prime}}^{\dag}|k(r^{\prime})\rangle . \nonumber
\end{eqnarray}
 The Green's function is defined here as
\begin{equation}
\label{greens-continuum}
g(\rra,\rrb;\varepsilon_{\kappa})= \frac{1}{W}\sum_{\kappa}\left\{\begin{array}[c]{cc}
|w_{\kappa}(r)\rangle\langle u^\ast_{\kappa}(r^{\prime})|&\,r>r^{\prime}\\
|u_{\kappa}(r)\rangle\langle w^\ast_{\kappa}(r^{\prime})|&\,r<r^{\prime}, \end{array}
\right.
\end{equation}
where the Wronskian $W$ acts as a renormalization factor. The two-dimensional spinors $|u_{\kappa}(r)\rangle$ and $|w_{\kappa}(r)\rangle$ are the
regular and irregular wavefunctions of an excited state $\kappa$.

The nonspectral method [Eq.~(\ref{greens-continuum})] has several advantages. First, for energies above the $\varepsilon_{\kappa}=0$ continuum limit,
the function $|w_{\kappa}(r)\rangle$ is a plane wave with complex values. In other words, for $E>\lambda$ (which accounts for all excitation energies
above the particle emission threshold) the strength function [Eq.~(\ref{strenghtfunction})] is always nonzero. This finite distribution of $S(E)$ is
a direct source of information for the escape width $\Gamma^{\uparrow}$ of the giant resonance \cite{DR.09}. Furthermore, all levels embedded in the
continuum and included in the Green's function [Eq.~(\ref{greens-continuum})] are characterized by the quantum number $\kappa$ and not by the
principal quantum number $n$, as done in the discrete approximation. In other words, there is no distinction between e.g. states $2p_{3/2}$ and
$3p_{3/2}$. In this way, the sum over a large number of unbound states is reduced to a sum of only a few $\kappa$ states. In addition, the excitations
to the Dirac sea (which includes the antiparticle states) are taken into account automatically and thus their calculation is redundant. Both these
characteristics can lead not only to a substantial reduction of the numerical effort but also to the absence of a corresponding cut-off parameter, as
required in the discrete RPA approach \cite{RMG.01}.

As discussed before, for open-shell nuclei, the BCS approximation is used at both the RMF and the RPA levels in order to take the pairing correlations
into account. As a consequence, the bound states in the pairing active space are described by an occupation factor, i.e, as quasiparticles. However,
since the continuum states above $E=0$ remain pure particle states, the bound and unbound states are not treated on an equal footing anymore. For this
reason, the response function of Eq.~(\ref{cont_term}) cannot be properly described within the quasiparticle CRPA model which needs to be extended to
account for the two-quasiparticle excitation in the pairing active space. This is done by adding the term
\begin{eqnarray}
\label{2qp_term}
\mathcal{R}_{2qp}^{0}(E) &=&
\sum_{k\leq
k^{\prime}}^{<E_{pair}}\frac{\eta_{kk^{\prime}}^{S+S^{\prime}}}{1+\delta_{kk^{\prime}}}
\langle k||Q_{c}||k^{\prime}\rangle_{r}\langle k||Q_{c^{\prime}}||k^{\prime}\rangle_{r^{\prime}} \nonumber\\
&&\times
\left(\frac{1}{\tilde{E}-E_{kk^{\prime}}}-\frac{1}{\tilde{E}+E_{kk^{\prime}}}\right) \nonumber\\
&+&
\sum_{k\leq k^{\prime}}^{<E_{p}}\frac{1}{1+\delta_{kk^{\prime}}}
\langle k||Q_{c}||k^{\prime}\rangle_{r}\langle k||Q_{c^{\prime}}||k^{\prime}\rangle_{r^{\prime}} \nonumber \\
&&
\times\left\{\frac{v_{k}^{2}}{\tilde{E}-\Omega_{k,k^{\prime}}}-\frac{v_{k}^{2}}{\tilde{E}+\Omega_{k,k^{\prime}}}\right.\nonumber\\ 
&&
\left.+\frac{v_{k^{\prime}}^{2}}{\tilde{E}-\Omega_{k^{\prime},k}}-\frac{v_{k^{\prime}}^{2}}{\tilde{E}+\Omega_{k^{\prime},k}}\right\},
\end{eqnarray}
where $E_{kk^{\prime}}=E_{k}+E_{k^{\prime}}$ is the two quasiparticle energy, $\tilde{E}=E+i\Gamma/2$ and $\Omega_{kk^{\prime}}=
E_{k}-\varepsilon_{k^{\prime}}-\lambda$. In Eq.~(\ref{2qp_term}), the first part corresponds to two-quasiparticle excitations, while the second part
is a correction term to prevent the excitations in the pairing active space from being counted twice (see Refs.~\cite{DR.11,HS.01} for more details).
Equation (\ref{2qp_term}) does not imply any significant additional numerical effort since the sum applies only in the pairing active space of a few
quasiparticle excitations. Consequently, a quasiparticle CRPA requires that the extended response function is properly treated only after the use of
\begin{equation}
\mathcal{R}^{0}(E) =\mathcal{R}_{cont}^{0}(E) +\mathcal{R}_{2qp}^{0}(E)
\end{equation}
 in Eq.~(\ref{reduced-bethe-salpeter}). The cross section for the photoabsorption processes is then derived from the strength function of
Eq.~(\ref{strenghtfunction}) according to
\begin{eqnarray}
 \sigma(E) &=& \frac{16\pi^{3}e^{2}}{9\hbar\,c}E\,S_{\text{RPA}}(E)\,\, [{\rm fm}^{2}] \nonumber \\
&=& 4.022\,E\,S_{\text{RPA}}(E)\,\, [{\rm mb}].
\end{eqnarray}

\section{Correction to the CRPA strength}
\label{CORRECT}
\subsection{Damping of the $E1$ strength}
\label{CONVOL}
Although the escape width $\Gamma^{\uparrow}$ is estimated microscopically within the CRPA, its value remains small, especially for heavy nuclei. This
is mainly due to the high centrifugal and Coulomb barriers, which prevent the excited nucleon from escaping. For that same reason, in light nuclei,
where the Coulomb and centrifugal barriers are smaller, the escape width $\Gamma^{\uparrow}$ dominates the total width \cite{DR.09}.

The remaining part of the total width corresponds to the coupling of the resonance to more complex configurations (2p-2h, 3p-3h, etc.) and its
contribution to heavy nuclei is large. One way to deal with the damping width is to calculate it in the second-RPA (SRPA) framework \cite{DNSW.90} or
by particle vibration coupling \cite{LPT.10,LRT.09,AGKK.11}.

Formally, self-energy insertions on particle and hole states spread the resonances and shift their centroids. In a first approximation, this can be
done by folding the RPA strength with a Lorentzian function
\begin{equation}
\label{lorentz}
 f_{L}(E,E^{\prime})=\frac{2}{\pi}\frac{\Gamma(E)E^{2}}{\left[E^{2}-(E^{\prime}-\Delta(E^{\prime})^{2}\right]^{2}+\Gamma(E)^{2}E^{2}}, 
\end{equation}
so that 
\begin{equation}
\label{convoluted_strength}
 S(E)=\int_{0}^{\infty}dE^{\prime}S_{\text{RPA}}(E^{\prime})f_{L}(E,E^{\prime}).
\end{equation}
Here $E^{\prime}$ is the excitation energy of the CRPA response, and $\Delta(E)$ and $\Gamma(E)$ the real and complex part of the self-energy,
respectively \cite{DNSW.90}. The energy-dependent width $\Gamma(E)$ can be calculated from the measured decay width of particle ($\gamma_{p}$) and
hole ($\gamma_{h}$) states \cite{DNSW.90}
\begin{equation}
\label{gamma}
 \Gamma(E)=\frac{1}{E}\int_{0}^{E}d\epsilon [\gamma_{p}(E)+\gamma_{h}(-E)](1+C_{ST}^{(\Gamma)}).
\end{equation}
This empirical way of determining $\Gamma(E)$ has the advantage of including, in principle, contributions from the excitation beyond 2p-2h
\cite{SW.88}. The resulting resonance width can be compared with experimental data, such as photoabsorption cross sections. Finally, the real part
$\Delta(E)$ of the self-energy is obtained from $\Gamma(E)$ by a dispersion relation \cite{DNSW.90}, where the interference factor $C_{ST}^{(E)}$ is
allowed to differ from $C_{ST}^{(\Gamma)}$ affecting the Lorentzian width of Eq.~(\ref{gamma}).

This method, though empirical, avoids important SRPA difficulties \cite{PR.10} related to the large deviation of the GDR energy and the need to
renormalize the residual interaction in order to deal with the spurious components \cite{GGC.10}.

\subsection{Comparison with the experimental GDR}

%
\begin{figure}[!b]
%
\centering
\includegraphics[width=230pt]{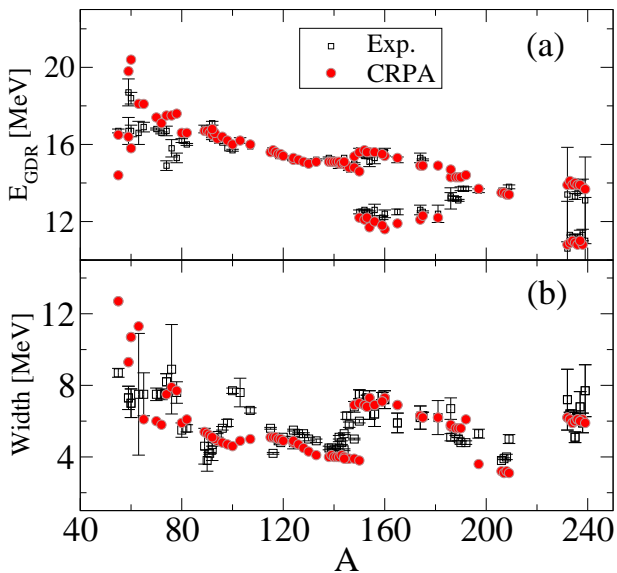}
\caption{(Color online) (a) Comparison between the experimental GDR energies for spherical and deformed nuclei (black squares) and the
GDR peak energies (red cycles) obtained after folding the CRPA distribution and including the deformation effects. (b) Similar comparison
between experimental and calculated GDR widths.}
\label{fig01}
\end{figure}

%
\begin{figure}[!t]
%
\centering
\includegraphics[width=250pt]{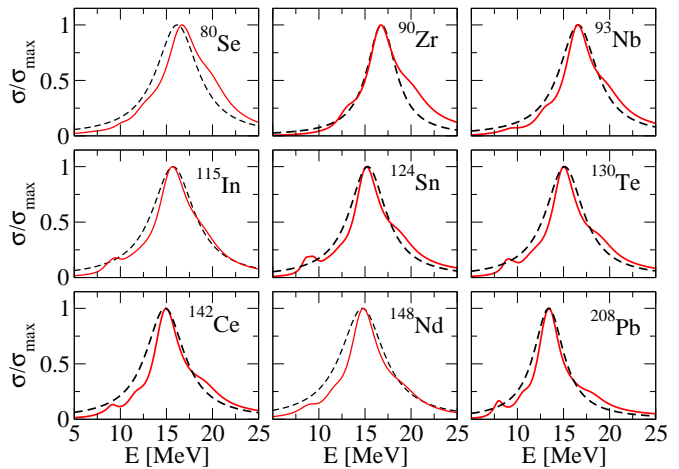}
\caption{(Color online) Comparison between the experimental photoabsorption cross section approximated by a simple Lorentzian curve (black dashed
line) and the CRPA predictions obtained with the PCF1 force (red solid line), for nine representative spherical nuclei given by ($Z,A$). All cross
sections are normalized to a peak cross section of unity. Experimental data are taken from \cite{ia00,RIPL03}}
\label{fig02}
\end{figure}

While the GDR of spherical nuclei can be well described by a Lorentzian-type function (Eq.~\ref{lorentz}), the situation for nuclei with a deformed
ground state is different. In particular, in the case of deformed nuclei with axial symmetry, the GDR is known to split into two major resonances
which correspond to oscillations of protons against neutrons along the principal axes, with frequencies inversely proportional to the length of these
axes \cite{PR.08}. In the phenomenological approach, the GDR peak energy is split according to the following rule \cite{Gaa.92}
\begin{eqnarray}
\label{split}
&E^{(1)}_{\text{GDR}}& + 2 E^{(2)}_{\text{GDR}} = 3 E_{\text{GDR}} \\
&E^{(2)}_{\text{GDR}}&/E^{(1)}_{\text{GDR}}=0.911\eta + 0.089, \nonumber
\end{eqnarray}
where $\eta$ is the ratio of the diameter along the axis of symmetry to the diameter along an axis perpendicular to it. Consequently, the folding
process of Eq.~(\ref{convoluted_strength}) takes place twice for each energy, with the two Lorentzians corresponding to the two energy peaks of
Eq.~(\ref{split}), leading to the final strength
\begin{equation}
\label{convoluted_strength2}
 S(E)=\int_{0}^{\infty}dE^{\prime}S_{\text{RPA}}(E^{\prime}) \left[\frac{1}{3}f^{(1)}_{L}(E,E^{\prime}) +
\frac{2}{3}f^{(2)}_{L}(E,E^{\prime})\right].\end{equation}
This procedure is applied to all nuclei, the parameter $\eta$ being derived from the HFB mass model of Ref.~\cite{GCP.10}.

%
\begin{figure}[!b]
%
\centering
\includegraphics[width=250pt]{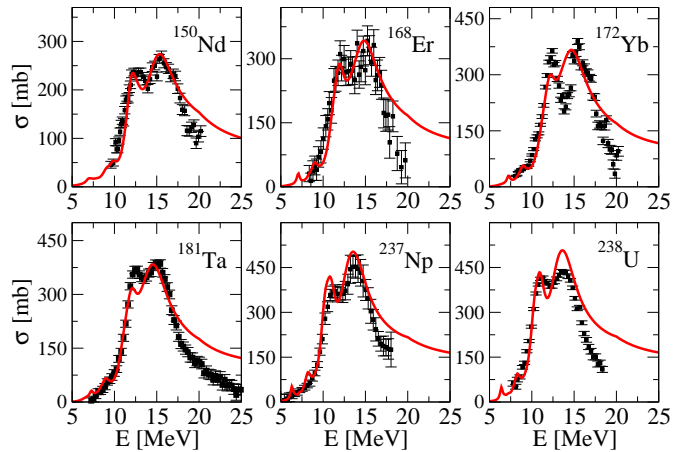}
\caption{(Color online) Comparison of the CRPA predictions (solid red line) with the experimental photoneutron cross sections (black dashed line) 
for several nuclei with well-deformed ground state. The experimental data are taken from \cite{CBB.71,GLM.81,GZ.76,BBV.68,Var.07,VBB.73} for the
nuclei $^{150}$Nd, $^{168}$Er, $^{172}$Yb, $^{181}$Ta, $^{237}$Np and $^{238}$U respectively.}
\label{fig03}
\end{figure}

The two interference factors $C_{ST}^{(E)}$ and $C_{ST}^{(\Gamma)}$ related to the energy shift and the resonance width of the $E1$ strength,
respectively, are calculated in a way similar to the parameter set of the Lagrangian, i.e. they are adjusted through a minimization procedure to
reproduce at best the measured GDR energy and width of experimentally known nuclei. In particular, the rms deviation factors
\begin{eqnarray}
 f^{(E)}_{\text{rms}} & = & 
\left[\frac{1}{N_{e}}\sum_{i=1}^{N_{e}}\left[E^{i}_{\text{GDR}}(th)-E^{i}_{\text{GDR}}(exp)\right]\right]^{1/2}
\end{eqnarray}
\begin{eqnarray}
 \nonumber \\
 f^{(\Gamma)}_{\text{rms}}&=&\left[\frac{1}{N_{e}}\sum_{i=1}^{N_{e}}\left[\Gamma^{i}_{\text{GDR}}(th)-\Gamma^{i}_{\text{GDR}}(exp)
\right]\right]^{1/2}\end{eqnarray}
are used for the GDR energies and the GDR widths, respectively. The best fit to experimental data is obtained for $C_{\text{ST}}^{(E)}=-0.85$ and
$C_{\text{ST}}^{(\Gamma)}=-0.59$, leading to $f^{(E)}_{\text{rms}} = 0.66$ ~MeV and $f^{(\Gamma)}_{\text{rms}} = 1.81$~MeV. This result is illustrated
in Fig.~\ref{fig01}, where the GDR energy and width are given for about 80 nuclei and compared with existing experimental measurements.

In Fig.~\ref{fig02}, we illustrate the $E1$ photoabsorption cross sections (normalized to unity) for several spherical nuclei all over the nuclear
chart. We see that globally, the GDR widths and energies are in good agreement with experimental data (represented by a Lorentzian function)
\cite{ia00,RIPL03}. When the deformation effect becomes important, it can be seen in Fig.~\ref{fig03} that the folded CRPA strength [corrected with
the prescription of Eq.~(\ref{split})] predicts rather well not only the GDR width, but also the splitting into the two observed peaks.

\subsection{Temperature dependence}
\label{Tdep}

%
\begin{figure}[!b]
%
\centering
\includegraphics[width=250pt]{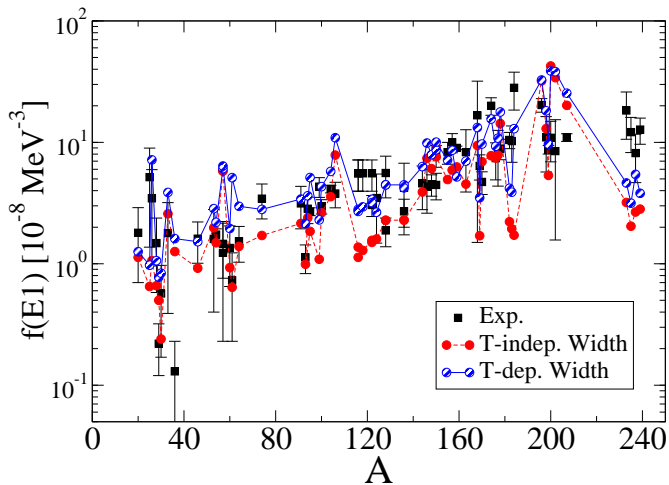}
\caption{(Color online) Comparison of the $T$-dependent and $T$-independent low-energy $E1$-strength functions deduced from the CRPA with the
experimental compilation \cite{RIPL03} including resolved-resonance and thermal-captures measurements, as well as photonuclear data for nuclei from
$^{25}$Mg up to $^{239}$U at energies ranging from 4 to 8 MeV.}
\label{fig04}
\end{figure}

%
\begin{figure}[!t]
%
\centering
\includegraphics[width=250pt]{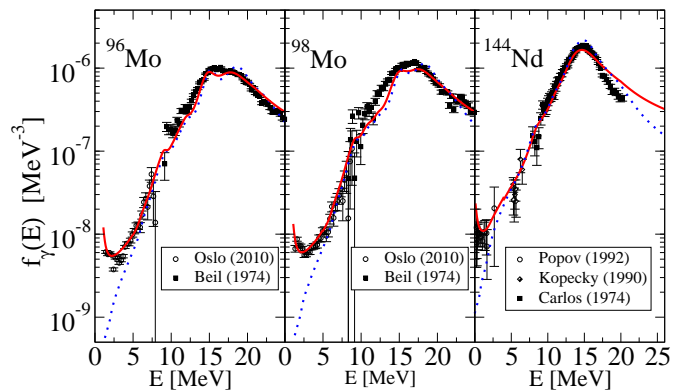}
\caption{(Color online) CRPA $\gamma$-ray strength of $^{96}$Mo, $^{98}$M and $^{144}$Nd, obtained with a $T$-independent (blue dotted line) and
$T$-dependent (red solid line) width. The results are compared with experimental data for the Mo \cite{Toft.10,BBC.74} and $^{144}$Nd isotopes
\cite{KU.90,Pop.82,Car.74}. The CRPA calculations corresponds to a temperature $T = 0.50~$MeV.}
\label{fig05}
\end{figure}

As already discussed, when dealing with practical applications and more precisely radiative captures, the deexcitation strength function needs to be
estimated. For low-energy incident particles, a reliable description of the tail of the GDR close to the particle threshold is required. 

The excitation of a nuclear state in this energy range is followed by a subsequent decay to the ground state which takes place via multiple
intermediate states of finite lifetime. The observation that in many nuclei the deexcitation $\gamma$-strength at $E\rightarrow 0$ is nonzero
\cite{LG.10} suggests that the level density in this region is high and that the $E1$ strength function depends on the nuclear temperature
\cite{KM.83,KU.90}. To provide a qualitative agreement with $\gamma$-decay measurements, a temperature dependence is traditionally introduced in the
expression of the GDR width \cite{KM.83,KU.90,LG.10}, i.e.,
\begin{equation}
\label{temperature}
\Gamma^{\prime}(E,T)=\Gamma(E)\frac{1}{E^{2}_{\text{GDR}}}\left[E^{2}+\frac{\alpha 4\pi^{2} T^{2}E_{\text{GDR}}}{E+\delta}\right],
\end{equation}
where $T$ is interpreted as the nuclear temperature of the final state and is estimated by $T_{f}\propto \sqrt{E_{f}}$. The constant parameter
$\delta=0.1$~MeV is introduced to ensure a nondivergent width at $E\rightarrow 0$ \cite{LG.10}.

Available data on thermal capture measurements from Mg to U have been used as a benchmark for the validity of the model and also as a way to derive
the parameter $\alpha$. In Fig.~\ref{fig04} the corresponding strength functions $f_{\gamma}(E)$ are shown at energies ranging between 4 and 8 MeV
depending on the corresponding binding energies. It is found that the value $\alpha=2.30$ gives the best overall agreement with the experimental
strength. We see that the temperature dependence improves the prediction, as compared to the $T=0$ strength.

The importance of the temperature dependence can be even more apparent for nuclei with enhanced $E1$ strength at low excitation energies, as for
instance in the deexcitation strength of $^{96}$Mo, $^{98}$Mo or $^{144}$Nd (Fig.~\ref{fig05}). After a proper renormalization of the position of the
GDR, the experimental strength at $E \sim 0$ is found to be well reproduced by the convoluted strength function with the temperature-dependent width
of Eq.~(\ref{gamma}). 

%
\begin{figure}[!t]
%
\centering
\includegraphics[width=230pt]{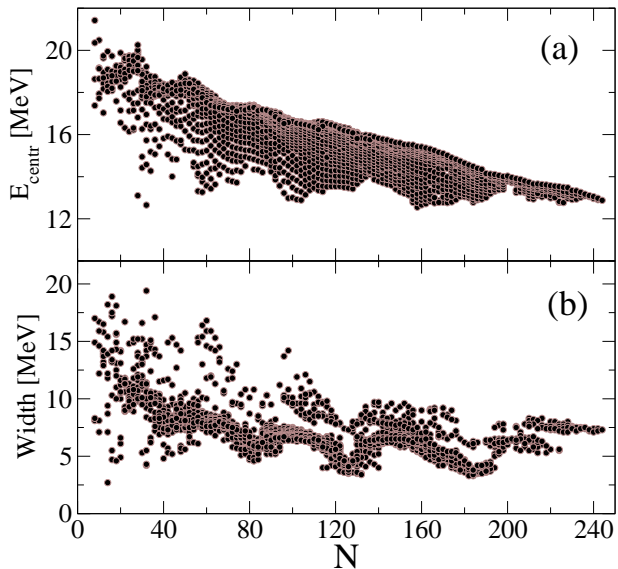}
\caption{(Color online) CRPA centroid energies (a) and widths (b) of even-even $A$ nuclei with $8\le Z\le 110$ and $10\le N\le 240$.}
\label{fig06}
\end{figure}

\section{Large scale calculations}
\label{HUGE}

Large-scale calculations based on the extended CRPA have been performed for all $8\le Z \le 110$ nuclei lying between the proton and neutron drip
lines, i.e. some 8000 nuclei. For all the even-even nuclei, the CRPA $E1$ distribution is folded by a Lorentzian-type function which properly takes
into account the effect of the interference factors, the deformation and the temperature dependence, as described in Sec.~\ref{CORRECT}. Interpolation
techniques are used to calculate the strength of the odd-$A$ and odd-odd nuclei. This interpolation is based on a simple linear averaging between the
two closest even-even nuclei, with respect to the GDR energies, i.e for an odd nucleus $(Z,N)$, the $E1$ strength is given by 
\[
S^{(N)}(E)=\frac{S^{(N-1)}(E+\lambda/2)+S^{(N+1)}(E-\lambda/2)}{2}
\]
where the quantity $\lambda=E_{\text{GDR}}^{(N-1)}-E_{\text{GDR}}^{(N+1)}$ corresponds to the difference in the GDR centroid energies between the
nuclei $(Z,N-1)$ and $(Z,N+1)$. Despite the empirical nature of this method, experimental results on low-lying $E1$ strength \cite{GCA.05} and GDR
data \cite{RIPL03} suggest that it can be adequately used in medium and heavy nuclei. A proper microscopic treatment of odd nuclei has been developed
\cite{Ka.01}, but remains numerically too heavy to be applied to large scale calculations.

%
\begin{figure}[!b]
%
\centering
\includegraphics[width=230pt]{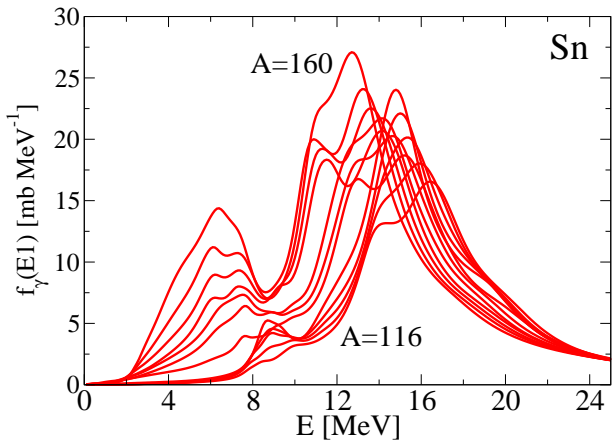}
\caption{(Color online) CRPA $E1$ strength function for the Sn isotopes with $A=116$ to $A=160$ by steps of 4 as a function of the energy.}
\label{fig07}
\end{figure}

%
\begin{figure}[!b]
%
\centering
\includegraphics[width=230pt]{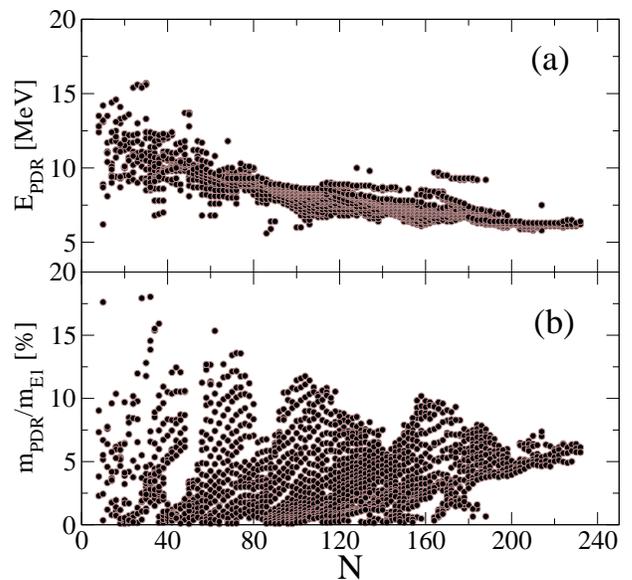}
\caption{(Color online) Same as Fig.~\ref{fig06} for the position of the PDR (a) and the ratio of PDR strength to the full $E1$ strength (b).}
\label{fig08}
\end{figure}

In Fig.~\ref{fig06}, the GDR centroid energies and widths are plotted with respect to the neutron number $N$ for all even-even nuclei. It can be
seen that the GDR energy decreases with the atomic mass while the GDR width reaches local minima in the vicinity of magic numbers, thus being
sensitive to shell effects. For extremely neutron-rich nuclei with $N>200$, the GDR width appears to remain rather constant around $8$~MeV.

Apart from the GDR properties which dominates the photoabsorption cross section, a second weaker mode appears at low energy, especially for nuclei
with a large neutron excess. This pygmy dipole resonance (PDR), which has been extensively studied in the last decade (see, e.g.,
\cite{Gor.98,Khan.02,DG.11}), is found at low energies, typically around 9~MeV for nuclei around the valley of $\beta$ stability and at even lower
energies for nuclei close to the drip lines (Fig.~\ref{fig07}). Since this excitation mode lies in the low-energy tail of the GDR, it is expected to
have a significant impact on the determination of the radiative neutron capture rates of exotic neutron-rich nuclei \cite{AR.07,Gor.98,Khan.02}.

In the upper panel of Fig.~\ref{fig08} the energy of the pygmy resonance is shown for all even-even nuclei up to $N=240$. We see that the PDR energy
decreases as we move to more neutron rich nuclei, a pattern similar to the one characterizing the GDR. This decrease is however slower and for
superheavy nuclei, $E_{\text{PDR}}$ remains rather constant. In Fig.~\ref{fig08}(b), the integrated PDR strength relative to the full $E1$
strength is given as a function of the neutron number $N$. The PDR strength is obtained by integrating over the energy region around the PDR peak
energy and up to about 10~MeV, where the GDR contribution dominates. The behavior of the PDR strength shown in Fig.~\ref{fig08} demonstrates the shell
effects affecting the pygmy mode; close to the neutron magic numbers $N=$20, 50, 82 and 126 a significant decrease of the PDR strength is observed,
while far from the magic numbers, this strength can become as large as 15\% of the total strength. Finally, for the neutron-rich nuclei with $N>200$,
the PDR strength remains constant at about 5\% of the GDR strength.

\section{Proton pygmy modes}
\label{PPDR}
The nature of the PDR in neutron-rich nuclei, as discussed in the previous section, refers to the well-accepted picture of a neutron skin vibration
with respect to a neutron-proton core.

In principle such a picture can be generalized to neutron-deficient nuclei for which an oscillation of the weakly bound proton skin could take place
against the isospin-saturated proton-neutron core. This idea is of course not new and has been studied in the past, though essentially on a
theoretical ground \cite{PVR2005,BCL.08}. 

%
\begin{figure}[!b]
%
\centering
\includegraphics[width=250pt]{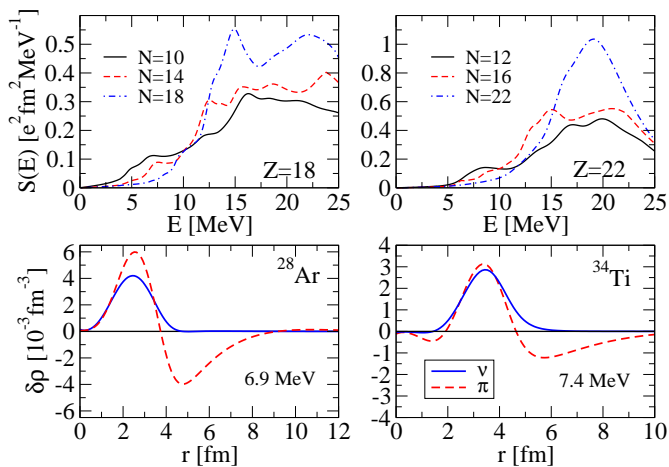}
\caption{(Color online) (Upper panels) $E1$ strength of the light Ar and Ti isotopes, showing a pronounced proton skin, as compared to the $N=Z$
isotopes (dot-dashed lines). (Lower panels) $^{28}$Ar (left) and $^{34}$Ti (right) proton and neutron transition densities at the PDR.}
\label{fig09}
\end{figure}

While many neutron-rich system exist, neutron-deficient nuclei are restricted to $Z<50$ elements, the Coulomb barrier preventing the creation of
heavier proton-rich systems. In light neutron-deficient nuclei, the proton excess remains rather small, so that a proton PDR is expected to be rather
weak, at least in comparison with the neutron PDR on the neutron-rich side of the valley of $\beta$ stability. In addition, the $E1$ strength in light
nuclei is quite spread, blending a possible proton PDR contribution.

%
\begin{figure}[!t]
%
\centering
\includegraphics[width=230pt]{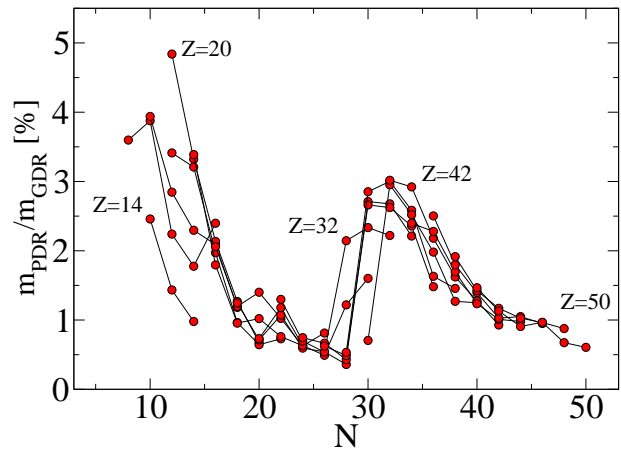}
\caption{(Color online) Ratio of the PDR to GDR strength with respect to the neutron number $N$ for the even-even proton-rich nuclei, from the proton
drip line up to the $N=Z$ isotopes}
\label{fig10}
\end{figure}

On the basis of our theoretical approach, we can identify and study the properties of the proton PDR. As shown in Fig.~\ref{fig09}, the
neutron-deficient $^{28}$Ar and $^{34}$Ti isotopes reveal an enhanced $E1$ strength around 6.9~MeV and 7.4~MeV, respectively, which could be
identified as a proton PDR. The first indication for this is that the low-lying strength disappears, as we reach the $N=Z$ isotopes, i.e. for
$^{36}$Ar and $^{44}$Ti. Furthermore, the proton and neutron transition densities (lower panels of Fig.~\ref{fig09}) clearly show no contribution of
any neutrons in the surface region; instead, we observe the existence of a proton skin which oscillates against the proton-neutron core, as suspected
in the first place.

In Fig.~\ref{fig10}, we estimate the proton PDR strength relative to the GDR strength for the even-even neutron-deficient nuclei with $Z\le 50$. The
quantity $m_{\rm PDR}$ corresponds to the integrated $E1$ strength in the 5-10~MeV region. For each isotopic chain, we observe a rapid decrease of
the PDR strength, as the mass number increases towards the $N=Z$ line or the magic numbers $N=20$, 28 or 50. This is a common behavior for all light
and medium nuclei, except for some specific chains around $Z\simeq 32$. This deviation from the general trend is also found for the neutron PDR
strength for nuclei around $N=50$, $N=82$ and $N\simeq 144$, as seen in Fig~\ref{fig08}.

\section{Radiative neutron capture rates}
\label{RATES}

The Maxwellian-averaged radiative neutron capture rates of astrophysical interest (see e.g the review \cite{AR.08} on the $r$-process nucleosynthesis)
are estimated within the statistical model of Hauser-Feshbach at a typical stellar temperature of $10^{9}$K, making use of the
$\footnotesize{\text{TALYS}}$ code 
\cite{KO.08,GO.08}. This version benefits in particular from an improved description of the nuclear ground state properties derived from the
nonrelativistic Hartree-Fock-Bogoliubov (HFB) method \cite{GO.10}, as well as from a nuclear level density prescription based on the combinatorial
model \cite{GO.08b}. The direct capture contribution as well as the possible overestimate of the statistical predictions for resonance-deficient
nuclei are effects that could have a significant impact on the radiative neutron captures by exotic nuclei \cite{Gor.98}, but are not included in the
present analysis. 
%
\begin{figure}[!b]
%
\centering
\includegraphics[width=220pt]{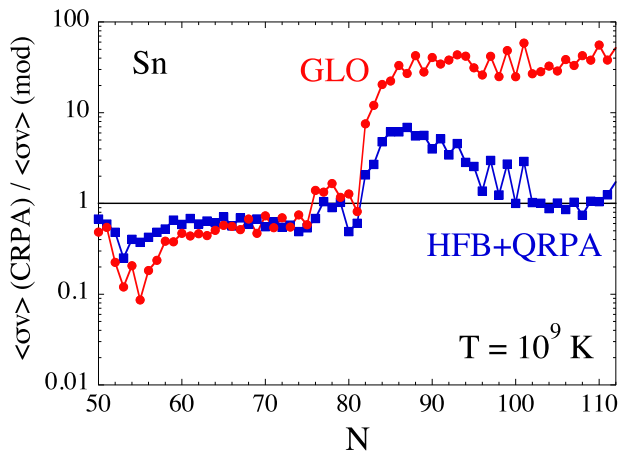}
\caption{(Color online) Ratio of the Maxwellian-averaged ($n$,$\gamma$) rates (at a temperature of $10^{9}$K) for the Sn isotopes obtained with the
CRPA $E1$ strength to the one using the HFB+QRPA \cite{Khan.04} (squares) or the GLO model~\cite{KU.90} (circles).}
\label{fig11}
\end{figure}

%
\begin{figure}[!t]
%
\centering
\includegraphics[width=220pt]{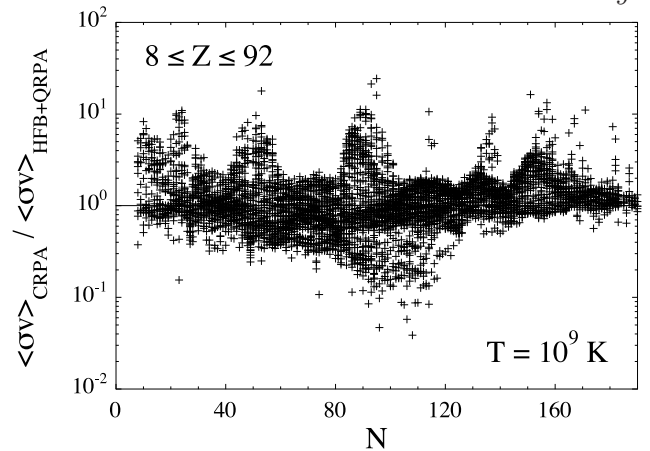}
\caption{(Color online) Ratio of the Maxwellian-averaged ($n$,$\gamma$) rate (at a temperature of $10^{9}$K) obtained with the CRPA $E1$ strength to
the one using the HFB+QRPA \cite{Khan.04} for all nuclei with $8\le Z \le 92$ lying between the proton and neutron drip lines.}
\label{fig12}
\end{figure}

In Fig.~\ref{fig11}, we compare the radiative neutron capture rates $\langle\sigma v\rangle$ for the Sn isotopes obtained with our $T$-dependent CRPA
with both the nonrelativistic HFB+QRPA predictions \cite{Khan.04} and the widely used phenomenological generalized Lorentzian model (GLO) of
Ref.~\cite{KU.90}. Although all three models lead to reaction rates within a factor of 2 for Sn nuclei below $N=82$, this is not the case for exotic
neutron-rich nuclei, for which the CRPA model gives rise to radiative neutron capture rates about 10 to 60 times larger than those obtained with the 
GLO model. This can be explained by the presence of a strong low-lying CRPA strength for neutron-rich nuclei, as shown in Figs.~\ref{fig07} and
\ref{fig08}.

A similar, though less pronounced, pattern is observed with respect to the nonrelativistic HFB+QRPA approach \cite{Khan.04}. In fact, in the region
between $^{132}$Sn and $^{142}$Sn, reaction rates larger by a a ratio of about 5 is found with our CRPA approach. This is due to the fact that the
neutron threshold for these nuclei lies at energies close to the pygmy resonance, which is generally more pronounced in the relativistic model. In
contrast, heavier isotopes, i.e. $^{146}$Sn or above have a neutron threshold below 3 MeV and are less affected by the PDR, which, although stronger,
lies at higher energies. For this reason, both models tend to give rise to similar reaction rates for the most exotic neutron-rich Sn isotopes.
Calculations based on a relativistic \cite{LLL.09} or nonrelativistic \cite{AGKK.11} QTBA description of the Sn isotopes show a similar behavior
when compared to the HFB+QRPA approach \cite{Khan.04}, although the low-lying strength obtained within the time-blocking approximation appears to be
more fragmented \cite{LPT.10,LRT.09,AGKK.11}.

The ratio of the CRPA neutron capture rates to those calculated with the nonrelativistic HFB+QRPA are shown in Fig.~\ref{fig12} for all nuclei with
$8\le Z \le 92$. Globally, the CRPA calculation tends to predict larger $E1$ strength at low energies mainly due to the lower energies at which the
PDR dominates. In some very neutron-rich nuclei the rates calculated with the CRPA strength can be about 10 times larger. Finally note that the
$T$-dependence introduced on the CRPA strength in Eq.~(\ref{temperature}) plays a minor role on the neutron capture rates of exotic neutron-rich
nuclei since their low neutron separation energy implies a low temperature of the deexciting levels.

\section{Conclusions}
\label{SUMMARY}

Starting from a point coupling Lagrangian with the parameter set PCF1 \cite{BMM.02}, we have used the relativistic continuum QRPA approach to study
the $E1$ collective excitation spectra of all nuclei between the neutron and proton drip lines. The RMF equations are solved in the coordinate space
self-consistently. For open-shell nuclei, the BCS model is applied to treat the pairing correlations.

The residual particle-hole interaction used in the RPA calculations is derived from the same Lagrangian in a fully self-consistent way. A set of two
interference factors $C_{ST}^{(E)}$ and $C_{ST}^{(\Gamma)}$ is additionally introduced in order to adjust systematically the damping width and the
centroid energy of the GDR energy on all photoabsorption data. In addition, the deformation and temperature effects are included to account for a
complete description of the nuclear dynamical problem. This extended CRPA model has been tested for nuclei for which experimental information is
available and a satisfactory quantitative agreement is found. In particular, the predicted GDR peak energies globally reproduce experimental data
within only a few hundred keVs.

On this basis, the extended CRPA model is used to perform a large-scale calculation of the $E1$ strength for all nuclei with $8\le Z \le 110$ lying
between the proton and neutron drip lines. As far as the collective dipole excitation is concerned, a clear pattern around magic numbers is observed
on the GDR width and the PDR strength, both being affected by the shell effects. The temperature dependence of the width also appears to affect the
strength in the low-energy region. 

For neutron-rich as well as light neutron-deficient nuclei, a low-lying PDR strength is found systematically in the 5--10~MeV region. The
corresponding strength can reach 10\% of the GDR strength in the neutron-rich region and about 5\% in the neutron-deficient region, and is found to
be significantly reduced in the vicinity of the shell closures. Finally, the neutron capture reaction rates of neutron-rich nuclei determined with the
CRPA strength can be about 2--5 times larger than those predicted on the basis of the nonrelativistic HFB+QRPA strength and up to about 50 times
larger than classically determined with the GLO model.

We have shown that the present approximation can described rather successfully the collective properties of stable nuclei for which experimental data
are available. However, for more exotic nuclei, in particular close to the drip lines, some of the phenomenological corrections used in the present
approach need to be replaced by sounder models. More specifically, an improved treatment of the pairing correlations can be achieved within the
Relativistic Hartree-Bogoliubov model \cite{Vretenar05} and the effects linked to the spreading of the GDR strength beyond the 1p-1h excitations as
well as the ground state deformation and odd number of particles need to be tackled following the recent works of \cite{DR.11,MPD.11,LRT.09,Ka.01}.


 \begin{acknowledgments}
Helpful discussions with P. Ring are gratefully acknowledged. This research has been supported by the FNRS (Belgium) and the Communaut\'e 
fran\c{c}aise de Belgique (Actions de Recherche Concert\'ees). 
 \end{acknowledgments}

\bibliographystyle{prsty} 

\end{document}